\documentclass[11pt,dvips]{article}
\usepackage{graphicx}
\begin{document}

\large \baselineskip4ex

\begin{center}
{\noindent \LARGE\bf Lattice vibrations and structural instability in Cesium near the
cubic to tetragonal transition
\\} \vspace{0.2in}
 {\Large Y. Kong and O. Jepsen\\ \vspace{0.3cm}{\normalsize {\em Max-Planck-Institut
f\"{u}r Festk\"{o}rperforschung, Heisenbergstr.1, D-70569 Stuttgart,
Germany\\}}  }

\vspace{0.3in}{\Large\bf Abstract}
\end{center}

Under pressure cesium undergoes a transition from a high-pressure fcc phase
(Cs-II) to a collapsed fcc phase (Cs-III) near 4.2GPa. At 4.4GPa there follows 
a transition to the tetragonal Cs-IV phase. In order to investigate the 
lattice vibrations in the fcc phase and seek a possible dynamical instability 
of the lattice, the phonon spectra of fcc-Cs at volumes near the 
III to IV transition are calculated using Savrasov's density 
functional linear-response LMTO method. Compared with quasiharmonic model 
calculations including non-central interatomic forces up to second 
neighbours, at the volume $V/V_0=0.44$ ($V_0$ is the experimental volume 
of bcc-Cs with $a_0$=6.048{\AA}), the linear-response calculations show 
soft intermediate wavelength $T_{[1\bar{1}0]}[{\xi}{\xi}0]$ phonons. 
Similar softening is also observed for short wavelength $L[\xi\xi\xi]$ and  
$L[00\xi]$ phonons and intermediate wavelength $L[\xi\xi\xi]$ phonons. 
The Born-von K\'{a}rm\'{a}n analysis of dispersion curves indicates that 
the interplanar force constants exhibit oscillating behaviours against 
plane spacing $n$ and the large softening of intermediate wavelength 
$T_{[1\bar{1}0]}[{\xi}{\xi}0]$ phonons results from a negative 
(110)-interplanar force-constant $\Phi_{n=2}$. The calculated frequencies 
for high-symmetry $K$ and $W$ phonons and longitudinal $X$ and $L$ phonons 
decrease with volume compression. In particular, the frequencies of the 
$T_{[1\bar{1}0]}[{\xi}{\xi}0]$ phonons with $\xi$ around $\frac{1}{3}$ 
become imaginary and the fcc structure becomes dynamically unstable for 
volumes below $0.41V_0$. It is suggested that superstructures corresponding 
to the $\mathbf{q}{\neq}0$ soft mode should be present as a precursor of 
tetragonal Cs-IV structure.

\vspace{0.5in} \noindent{\bf Keywords:} Cesium, Phonon, Linear-response
LMTO

\newpage
\section{Introduction}
At ambient pressure cesium metal crystallises in a bcc structure with
experimental lattice parameter $a_0=6.048\AA$\cite{anderson} at 0K, obtained
from the high-temperature values by extrapolation. It 
is a simple $s^1$ metal with nearly free-electron character. Under pressure 
the $s$ valence electrons are transferred to more localised $d$-like 
states\cite{sternheimer}, and cesium exhibits an interesting sequence of phase
transitions[3-8]. At the pressure 2.3GPa a transition\cite{kennedy,hall} from
bcc (Cs-I) to fcc (Cs-II) occurs with a small reduction of volume. Near
4.2GPa, the fcc Cs-II phase undergoes an isostructural
transition\cite{hall} to a collapsed fcc phase Cs-III with a large
volume reduction (9\%). Then at 4.4GPa there follows a transition to the
tetragonal Cs-IV phase\cite{takamura} with a decrease of the coordination
number from 12 in Cs-III to 8 in Cs-IV (strictly, 4 nearest neighbours at 
3.349\AA and 4 second-nearest neighbours at 3.542\AA at 8 GPa). At 
higher pressures more complicated phases\cite{takamura1,schwarz,takemura} 
appear.

\par Many detailed investigations have been devoted to the study of the
phase transitions in cesium, both experimentally (see Ref.\cite{schwarz}
and literature cited therein) and theoretically (see, e.g., Refs.[7-12]).
The pressure-induced electronic $s{\rightarrow}d$ transition was believed
to be the driving force for destabilising the highly symmetric
low-pressure structures (bcc or fcc) with respect to lower symmetry
structures. The transition III$\rightarrow$IV was attributed to the
accelerated progress\cite{takamura} of the $s{\rightarrow}d$ transition
and the unusual decrease of the coordination number from Cs-III to Cs-IV
has been interpreted in terms of directional bonding induced by the
$s{\rightarrow}d$ transition\cite{mcmahan1,andersen}.

\par In general, the $s{\rightarrow}d$ electronic transition in
cesium affects not only the static lattice properties by changing the
electronic band structure but also the lattice vibrational properties by
modifying the effective interatomic interactions.  
The structural behaviour of cesium may therefore be reflected in the phonon
spectrum and it is therefore of interest to investigate the phononic behaviour 
with applied pressure. Gl\"{o}tzel
and McMahan\cite{gloetzel} suggested that an anomaly in the lattice
vibrational contribution to the pressure is a possible mechanism for the
isostructural II$\rightarrow$III transition. Recently, Christensen 
et al.\cite{christensen} calculated phonon dispersions of fcc-Cs within 
the quasiharmonic approximation to study the thermal expansion coefficient 
of cesium and the isostructural transition. They found that below 
$V/V_0=0.375$ ($V_0$ is the experimental volume of bcc-Cs at ambient 
conditions) fcc-Cs becomes unstable due to softening of a transverse [110] 
phonon branch, which is characterised by a negative shear elastic constant
$C^{\prime}$. Very recently, Xie et al.\cite{xie} investigated the phonon
instabilities of cesium using density-functional perturbation theory. Their
calculations showed that the instability of the fcc cesium occurs at a volume
of about 0.46$V_0$, which is also driven by a negative $C^{\prime}$. However, 
it is not clear what happened with the phonons before the $C^{\prime}$ goes 
soft. Using the density functional linear-response LMTO method\cite{savrasov}, 
we calculate the accurate phonon spectrum of fcc-Cs at the volumes near the 
III-IV transition to study the lattice vibrations in fcc-Cs and to seek a 
possible dynamical instability of fcc-Cs before $C^{\prime}$ becomes negative. 
According to our calculations, fcc-Cs becomes unstable at a volume slightly 
smaller than $V/V_0=0.41$. Soft $T_{[1\bar{1}0]}$-$[{\xi\xi}0]$ phonons 
with $\xi\sim\frac{1}{3}$ are found to be responsible for the dynamical
instability.

\par This paper is organised as follows. Some computational details concerning
linear-response LMTO calculations are described in the next section. Section 3
presents the results obtained and some discussions. Firstly, the phonon 
dispersion curves for fcc-Cs at the volume $V/V_0=0.44$ are analysed in terms 
of force constants between atoms; then the phonon frequencies at 
various volumes are discussed in relation to the instability of the fcc 
structure. Several concluding remarks are given in the last section.

\section{Computational Details}

\par The density functional linear-response LMTO method\cite{savrasov} is 
used in the present study to calculate the phonon spectrum of fcc-Cs at the 
volumes around the Cs-III - Cs-IV transition. Phonon dispersion curves for 
a large number of simple and transition metals and a few compounds have been 
calculated by the linear-response LMTO method, and excellent agreement 
with experimental data were obtained.\cite{savrasov,savrasov1} In the 
following we summarise some details of the calculations.

\par The dynamical matrix of fcc-Cs is calculated for a set of irreducible
{\bf q} points in a (8,8,8)-reciprocal lattice grid (29 points per 1/48th 
part of the Brillouin Zone (BZ)). The (I,J,K) reciprocal-lattice grid is 
defined in the usual manner: $\mathbf{q}_{ijk}=(i/I)\mathbf{G}_1+
(j/J)\mathbf{G}_2+(k/K)\mathbf{G}_3$, where $\mathbf{G}_i$ are the primitive 
translations in reciprocal space. We chose the exchange-correlation potential 
of Vosko-Wilk-Nusair\cite{VWN} plus the non-local generalised-gradient-approximation 
(GGA-96) correction\cite{gga96}. Because this gave a better prediction of the 
equilibrium volume at ambient pressure and 0K which is only 4\% smaller than 
the experimental zero-pressure volume at 0K. On the other hand, the local 
density approximation (LDA) gave a nearly 20\% overbinding at ambient pressure. 
The calculated bcc to fcc transition pressure of 2.0GPa is in good agreement 
with the experimental value of 2.3GPa.\cite{hall}

In the calculations a $3{\kappa}-spd$ LMTO basis set is used with the
one-center expansions inside the non-overlapping muffin-tin (MT) spheres 
performed up to $l_{max}=6$. In the interstitial region, the $s$-, $p$- and 
$d$-basis functions are expanded in plane waves. The $5s$ semicore state is 
treated as valence state in a separate energy window. The induced charge 
densities and screened potentials are represented inside the MT spheres by 
spherical harmonics up to $l_{max}=6$ and by plane waves with a 148.4 Ry 
energy cutoff (9984 plane waves) in the interstitial region.

\par The $\mathbf{k}$-space integration is performed over a (16,16,16) grid 
(145 irreducible point) by means of the improved tetrahedron method\cite{tetra}, 
but the integration weights for these $\mathbf{k}$ points are 
calculated from a denser (32,32,32) grid (897 points in the irreducible BZ). 
This results in a more accurate representation of the Fermi surface 
with a small number of $\mathbf{k}$ points. 

\par The phonons along the high-symmetry
lines presented in the next section were calculated in a denser $\mathbf{q}$ 
mesh which fit to the (16,16,16) $\mathbf{k}$-grid.

\section{Results and Discussions}

\subsection{Phonon dispersions at $V/V_0=0.44$}

\par Fig. 1 shows the calculated phonon dispersion curves along some
high-symmetry directions for fcc-Cs at the volume 0.44$V_0$, which is  
in the experimental volume range of the cubic to tetragonal phase 
transition. To the right the calculated phonon density of states (DOS) 
is plotted. The dash lines represent the fitted results by the Born-von 
K\'{a}rm\'{a}n model which is described in the next section. The 
calculated phonon frequencies at the high-symmetry zone-boundary (ZB) 
points $L, X, K$ and $W$ are listed in Table 1.

\par As observed in Fig. 1, the phonon frequencies of the $L[{\xi}{\xi}0]$,
$T[00\xi]$ and $T[\xi\xi\xi]$ branches exhibit rather normal dispersion
while the $T[{\xi}{\xi}0]$, $L[00\xi]$ and
$L[\xi\xi\xi]$ dispersion curves show abnormal behaviour for intermediate 
and short wavelengths, which indicates long-range interactions between atoms. 
For the $L[\xi\xi\xi]$ phonon branch, small downwards anomaly is seen around 
$\xi=\frac{1}{4}$. Near the zone boundary, the $L[\xi\xi\xi]$ phonons have very
little dispersion and the frequencies of the $L[00\xi]$ phonons are rather 
small compared to other fcc metals. The $T[{\xi}{\xi}0]$ phonons have flat 
regions around ${\xi}=\frac{1}{3}$ and $\frac{3}{4}$ for the polarisation 
$[1\bar{1}0]$ ($T_1$) and $[001]$ ($T_2$), respectively. Since the phonon 
DOS in Fig. 1 was calculated at a coarse (8,8,8) grid, these abnormal 
behaviours are not clearly seen in the plotted DOS. Nevertheless, the 
sudden increase of DOS at the frequency just above 0.3 THz and the small 
sharp peak at the frequency near 1.6 THz indicate the presence of these 
anomalies.

\par In metals at equilibrium the long-range interatomic interactions 
are generally of little importance due to screening effect of the electrons. 
The lattice dynamical properties of metals can therefore be described well 
using a short-range force model. This, however, may not hold for cesium under 
high pressure. The pressure-induced electronic $s{\rightarrow}d$ transition 
causes the valence charge density of cesium to deviate from spherical symmetry 
with a consequent reduction in electronic screening and increased interactions
between the atoms. To check this, we calculate the phonon dispersion curves of 
fcc-Cs at the volume $V/V_0=0.44$ using a short-range force model within the 
quasiharmonic approximation. The dispersion relations are derived as in the 
usual harmonic approximation by considering non-central forces between atoms 
up to second neighbours and the four interatomic force constants involved in 
the dynamical matrices are obtained from the three calculated elastic 
constants\cite{cij} and the $T[100]$ 
zone boundary phonon frequency by expanding the elements of the dynamical
matrix and finding its long-wavelength limit. The obtained dispersion
curves are plotted in Fig. 1 by solid lines and the phonon frequencies
at the zone boundaries are listed in Table 1. Note that the quasiharmonic
results presented here are derived from a short-range force model,
which does not exactly consider the electronic contribution to the dynamical 
matrix. The difference to the results from linear-response calculations may 
therefore reveal the effects of long-range interatomic interactions. It may 
be seen in Fig. 1 that the short-range non-central force model describes the 
$T[00\xi]$, $T[\xi\xi\xi]$ and $L[{\xi\xi}0]$ phonon dispersions quite well. 
However, the previously mentioned anomalies in the $L[00\xi]$, $L[\xi\xi\xi]$ 
and $T[{\xi\xi}0]$ dispersion curves calculated by the linear-response method 
are not reproduced and they may therefore be a consequence of long-range
interactions. 

\par In general, an abnormal behaviour of the phonon dispersion may
be attributed to special properties of interatomic force constants. In the 
following subsection, we shall analyse the dispersion curves by using the 
Born-von K\'{a}rm\'{a}n model to find these special properties of the 
interactions between the atoms in fcc-Cs.

\subsection{Born-von K\'{a}rm\'{a}n analysis of force constants}

\par Since one of the motivations of the present work is to interpret the
anomalies in the calculated phonon dispersion curves for fcc-Cs, it is
necessary to use a model which can reproduce these anomalies. This could
be achieved by a force-constant model based on the Born-von K\'{a}rm\'{a}n
theory\cite{born}. 

\par In the Born-von K\'{a}rm\'{a}n theory the frequencies $\nu(\mathbf{q})$ 
of the normal vibration modes for an fcc lattice with one atom per unit cell 
are given as the solutions of a $3{\times}3$ determinantal equation
\begin{equation}\label{deter}
|4{\pi}^2M{\nu}^2(\mathbf{q})\delta_{\alpha\beta}-C_{\alpha\beta}(\mathbf{q})|=0,
\end{equation}
where $M$ is the mass of the atom and
\begin{equation}\label{calbe}
C_{\alpha\beta}(\mathbf{q})=\sum_{l^{\prime}}\Phi_{\alpha\beta}(l,l^{\prime})
exp[i\mathbf{q}\cdot\mathbf{R}(l,l^{\prime})],
\end{equation}
where $\Phi_{\alpha\beta}(l,l^{\prime})$ is the $\alpha\beta$ component of
the force constant matrix between the atoms in the $l$th and
$l^{\prime}$th cell. For the phonon modes along the three high-symmetry
directions $[{\xi\xi}0], [00\xi]$ and $[\xi\xi\xi]$, the $C_{\alpha\beta}$
matrices are diagonal and the solution of
Eq.(\ref{deter}) leads to equations in $\nu^2$ of the form
\begin{equation}\label{planar}
4{\pi}^2M{\nu}^2=\sum_{n}\Phi_n[1-cos(n{\pi}q/q_{max})],
\end{equation}
where $q_{max}$ is half the distance to the nearest reciprocal lattice
point in the direction of $\mathbf{q}$, $q=|\mathbf{q}|$ and $\Phi_n$ is a
sum of $\Phi_{\alpha\beta}(l,l^{\prime})$\cite{brockhause}
for which the phase $\mathbf{q}\cdot\mathbf{R}(l,l^{\prime})$ is a
constant. $\Phi_n$ effectively represents a force between a plane of atoms
and the planes of atoms normal to $\mathbf{q}$ and $n$ planes away.
The summation includes $N$ terms so that $\Phi_n=0$ for $n>N$. Thus,
a Fourier series analysis of the squares of the phonon frequencies 
will yield the interplanar force constants.

\par Accordingly, a least-squares Fourier fit are performed using 
Eq.(\ref{planar}) for the phonon dispersion curves of fcc-Cs at the volume 
0.44$V_0$ calculated from linear-response calculations. The fitted dispersion
curves are plotted in Fig. 1 by the dashed lines.  Within the accuracy of the
calculated phonon frequencies, it is found that a satisfied fit can 
only be obtained by including at least four planes in the $[00\xi]$ branches, five
planes in the $T[\xi\xi\xi]$ branch and even six planes in the $[{\xi\xi}0]$ and
$L[\xi\xi\xi]$ branches. In Table 2 we list the fitted interplanar force
constants $\Phi_n$. Since four planes 
along the $[00\xi]$ direction correspond to interactions out to at least eighth
neighbour atoms and five (six) planes along the $[\xi\xi\xi]$ ($[{\xi\xi}0]$)
direction out to farther neighbours, the fitted results confirm the
presence of long-range interactions between atoms in fcc-Cs. We plot
the fitted $\Phi_n$ for the $T_{[1\bar{1}0]}$[$\xi\xi$0] branch in Fig. 2a
as a function of $n$, which effectively is the distance between planes of atoms.
The $\Phi_n$ are observed to oscillate with the distance between planes,
especially with a prominently negative $\Phi_{n=2}$. The values of $\nu^2$
for the same phonon branch and the fitted curve with $N=8$ are shown in
Fig. 2b together with the individual contributions from the first five
Fourier components. It is seen that the $n=2$ planes produce a largely
negative contribution to the frequencies of phonons, which partly cancels
the dominant contribution from the $n=1$ planes. Consequently, the soft
intermediate-wavelength $T_{[1\bar{1}0]}[{\xi}{\xi}0]$ phonons result
and a flat region of the dispersion curve is formed. A similar oscillating
behaviour of the fitted $\Phi_n$ is seen in some of the other branches
although they are less pronounced. The anomalies observed in the
$T_{[001]}[{\xi}{\xi}0]$ and $L[{\xi}{\xi}\xi]$ dispersion curves can be
associated with these oscillations. For the $L[00\xi]$ branch no oscillations 
in $\Phi_n$ is found. The very low phonon frequencies for long-wavelengths 
compared to the short-range model calculation may be attributed to the small 
value of $\Phi_{n=1}$.

\par The interplanar force constants can be
used to derive interatomic force constants\cite{brockhause}. With a
general force model, a linear least-square fitting analysis for the
$\Phi_n$ in Table 2 only allows us to work out the interatomic force
constants in fcc-Cs extending to the fourth neighbour. The truncation of
the interactions with farther neighbours introduces large errors in the
fitting procedure and makes the fitted interatomic force constants less
reliable. Nevertheless, in Table 3 we list the best fit values of the
interatomic constants in fcc cesium.

\subsection{Soft mode and lattice instability}

\par The volume dependence of the high-symmetry zone-boundary $L, X, K$, and
$W$ phonons for fcc-Cs are presented in Fig. 3. From this the mode-Gr\"{u}neisen 
parameter $\gamma_i=-(\partial ln\nu_i/{\partial}lnV)$ was derived and listed 
in Table 1. It may be seen that all the high-symmetry phonons 
except the transverse $X$ and $L$ phonons decrease with increasing pressure
and consequently have negative mode-Gr\"{u}neisen parameters, which 
may be an indication that the fcc structure is unstable under pressure.

\par In Fig. 4 we show the $T_{[1\bar{1}0]}$[$\xi\xi$0]
phonon-dispersion curves of fcc-Cs for the volumes $V/V_0=0.37, 0.40,
0.41$, and $0.44$. At volumes between $V/V_0=0.41$ and 0.40 the phonon
frequencies around $\xi=\frac{1}{3}$ become imaginary. However, the shear 
elastic constant $C^{\prime}$ does not become negative before the volume is 
reduced to $V/V_0=0.37$ where the entire phonon branch becomes unstable.  
Christensen et al.\cite{christensen} and Xie et al.\cite{xie} inferred from 
their calculations that it is the pressure-induced negativity of $C^{\prime}$ 
which is the cause of the instability of the fcc phase. They however did not 
trace the development of the instability in details and it is clear from Fig. 
4 that it is a $\mathbf{q}{\neq}0$ $T_{[1\bar{1}0]}$[$\xi\xi$0] soft mode 
which drives the dynamical instability of the fcc phase close to $V/V_0=0.41$. 
As a consequence of the soft mode, superstructures should exist as a precursor 
to the tetragonal Cs-IV phase transition. It is interesting to note that 
the calculated transition volume $V/V_0=0.41\sim 0.40$ is close to the volume
where the fcc to tetragonal transition is observed\cite{takamura}. 
In Fig. 5 we illustrate the displacement pattern of atoms in fcc-Cs 
corresponding a $T_{[1\bar{1}0]}$[$\frac{1}{4}\frac{1}{4}$0] phonon mode 
and the resulting Cs superstructure. As seen from Fig. 5, a structural 
element of the displacement pattern is triangular prisms, which are also 
present in the Cs-IV structure, even though other possible soft modes as 
well as the coupling between phonons could be important. 

\par Usually a soft phonon mode leads to a second-order or nearly first-order 
phase transition. Early experimental results\cite{takamura} indicated 
that the Cs-III-Cs-IV transition is first-order in character with a 
4.3\% volume reduction. Therefore soft phonon modes should not be
considered the unique driving force in the cesium III-IV phase transformation 
even though the soft mode leads to an instability in the fcc structure.
The changes of the electronic properties under pressure, such as $s-d$ 
transition, must play an important role.

\section{Conclusions}
Using a density functional linear-response LMTO method, we have studied the
phonon spectra of fcc cesium at volumes near the III-IV transition.
Several anomalies in the dispersion curves have been found.

\par Within the Born-von K\'{a}rm\'{a}n theory of lattice dynamics, the
observed anomalies in the calculated phonon dispersion curves for fcc-Cs
at the volume 0.44$V_0$ are connected with oscillating behaviours of the
interplanar force constants against plane spacing $n$. In particular, a
negative (110)-interplanar force-constant $\Phi_{n=2}$ is found to be
responsible for soft intermediate-wavelength
$T_{[1\bar{1}0]}[{\xi}{\xi}0]$ phonons.

\par The frequencies of high-symmetry $K$ and $W$ and longitudinal $X$ 
and $L$ phonons decrease with volume compression and the consequent negative 
mode-Gr\"uneisen parameters indicate an instability of fcc lattice under 
pressure. When the volume is below $0.41V_0$, the frequencies of the 
$T_{[1\bar{1}0]}[{\xi}{\xi}0]$ phonons with $\xi$ around $\frac{1}{3}$
becomes imaginary prior to the sign change of the shear elastic constant
$C^{\prime}$ and the fcc lattice is thus dynamically unstable. It is 
suggested that superstructures corresponding to the $\mathbf{q}{\neq}0$
$T_{[1\bar{1}0]}[{\xi}{\xi}0]$ soft phonon should be present as a precursor 
of tetragonal Cs-IV structure. It was pointed out in Ref.\cite{schwarz} that
the fcc structure of Cs-III could not be confirmed and that Cs-III appears to
be a rather complicated structure. The present argument may stimulate 
further experimental study for reexamining the structure of Cs at the 
pressure near the Cs-II-Cs-III-Cs-IV transitions.

\section*{Acknowledgements}
Y. Kong gratefully acknowledge helpful discussions and advice from Prof. O. K. Andersen,
and we are grateful to Prof. K. Syassen for a critical reading of the manuscript.

\newpage \large
\section*{Tables}
~\\

\begin{table}[h]\caption{\label{phonon}Calculated phonon frequencies
$\nu_i$ (THz) at the high-symmetry {\sl L, X, K} and {\sl W} points for
fcc Cs at the volume $V/V_0=0.44$ and corresponding mode-Gr\"{u}neisen
parameters $\gamma_i$ defined by $\gamma_i=-(\partial ln\nu_i/\partial
lnV)$. Also listed in brackets are the frequencies from the short-range
force model calculation.}
\begin{center}
  \begin{tabular}{llrrrr}\hline \hline
    & &$L~~$&$X~~$&$K~~$ &$W~~$  \\ \hline
{\sl L~}-branch&$\nu_i$ &1.97 (2.13)&1.73 (2.10)&1.58 (1.65)&1.51 \\ \cline{2-6}
         &  $\gamma_i$&$-0.95$&$-0.79$&$-0.80$&$-1.27$ \\ \hline
{\sl T$_1$}-branch&$\nu_i$&0.82 (0.86)&1.36 (1.36)&0.95 (1.20)&1.23 \\ \cline{2-6}
         &  $\gamma_i$&0.71&0.49&$-0.80$&$-1.60$  \\ \hline
{\sl T$_2$}-branch&$\nu_i$&0.82 (0.86)&1.36 (1.36)&1.62 (1.93)&1.46  \\ \cline{2-6}
         &  $\gamma_i$&0.71&0.49&$-0.93$&$-1.41$ \\ \hline
  \end{tabular}
{}\end{center}
\end{table}

\vspace{0.5in}
\begin{table}[h]\caption{\label{planeforce}Interplanar force constants
$\Phi_n$ (up to $n=6$) for fcc Cs at the volume $V/V_0=0.44$ in units of
10$^{-3}$ Ry/bohr$^2$. The errors given are approximate errors for the fitted
$\Phi_n$. Note that a satisfactory fit for the
[00$\xi$] and $T$[$\xi\xi\xi$] phonons is obtained with
only four and five Fourier components, respectively.}
\begin{center}
  \begin{tabular}{rlrrrrrrr}\hline \hline
    & &$\Phi_1$&$\Phi_2$&$\Phi_3$&$\Phi_4$&$\Phi_5$&$\Phi_6$&error  \\ \hline
$T$[00$\xi$] & &$10.69$&$0.11$&$-0.24$&$-0.18$&$-~~$&$-~~$& $\pm$0.15
\\ $L${[}00$\xi$] & &$16.09$&$1.85$&$~0.72$&$~0.29$&$-~~$&$-~~$& $\pm$0.50
\\ \hline $T$[$\xi\xi\xi$] &
&$~4.11$&$~0.49$&$-0.36$&$~0.10$&$-0.05$&$-~~$& $\pm$0.02 \\
$L${[}$\xi\xi\xi$] & &$23.09$&$-0.86$&$-2.44$&$~3.35$&$~1.39$&$-2.78$&
$\pm$0.50 \\ \hline $T_{[001]}$[$\xi\xi$0] &
&$16.32$&$~3.56$&$-0.40$&$-1.06$&$~0.80$&$-0.55$&$\pm$0.19 \\
$T_{[1\bar{1}0]}${[}$\xi\xi$0] &
&$~8.51$&$-4.10$&$~1.74$&$-0.42$&$~0.21$&$~0.09$&$\pm$0.11
\\ $L${[}$\xi\xi$0] &
&$~8.88$&$14.52$&$~1.29$&$-0.85$&$~0.30$&$-0.29$&$\pm$0.12 \\ \hline
  \end{tabular}
{}\end{center}
\end{table}

\begin{table}[h]\caption{\label{force}The best fit values of interatomic
force-constants for fcc Cs at the volume $V/V_0=0.44$ in units of
Ry/bohr$^2$. The force constant notation follows the definition of
Brockhouse et al.\cite{brockhause}.}
\begin{center}
  \begin{tabular}{ccl}\hline \hline
Neighbour location&Force constants&Values (Ry/bohr$^2$) \\ \hline
\begin{tabular}{c}First \\ $\frac{a}{2}$(1,1,0) \\ \end{tabular} &
\begin{tabular}{ccc}$\alpha_1$&$\gamma_1$& 0 \\
$\gamma_1$&$\alpha_1$& 0 \\ 0&0&$\beta_1$ \end{tabular} &
\begin{tabular}{l}$\alpha_1=21.2{\times}10^{-4}$ \\ $\beta_1=-2.0{\times}10^{-4}$ \\
          $\gamma_1=38.6{\times}10^{-4}$ \end{tabular} \\ \hline
\begin{tabular}{c}Second \\ $\frac{a}{2}$(2,0,0) \\ \end{tabular} &
\begin{tabular}{ccc}$\alpha_2$& 0 & 0 \\ 0&$\beta_2$& 0 \\ 0&0&$\beta_2$ \end{tabular} &
\begin{tabular}{l}$\alpha_2=-2.1{\times}10^{-4}$ \\ $\beta_2=~5.8{\times}10^{-4}$ \end{tabular} \\ \hline
\begin{tabular}{c}Third \\ $\frac{a}{2}$(2,1,1) \\ \end{tabular} &
\begin{tabular}{ccc}$\alpha_3$&$\delta_3$&$\gamma_3$ \\ $\delta_3$&$\beta_3$&$\gamma_3$ \\
                   $\gamma_3$&$\gamma_3$&$\beta_3$ \end{tabular} &
\begin{tabular}{l}$\alpha_3=~3.6{\times}10^{-4}$ \\ $\beta_3=-0.6{\times}10^{-4}$ \\
      $\gamma_3=~3.6{\times}10^{-4}$ \\ $\delta_3=-0.4{\times}10^{-4}$ \end{tabular} \\ \hline
\begin{tabular}{c}Fourth \\ $\frac{a}{2}$(2,2,0) \\ \end{tabular} &
\begin{tabular}{ccc}$\alpha_4$&$\gamma_4$& 0 \\ $\gamma_4$&$\alpha_4$& 0 \\
                    0&0&$\beta_4$ \end{tabular} &
\begin{tabular}{l}$\alpha_4=-0.4{\times}10^{-4}$ \\ $\beta_4=0.04{\times}10^{-4}$ \\
          $\gamma_4=-4.1{\times}10^{-4}$ \end{tabular} \\ \hline
  \end{tabular}
{}\end{center}
\end{table}

\clearpage
\baselineskip4ex
\section*{Figure Captions}

\noindent Figure 1: Calculated phonon-dispersion curves along three symmetry
directions for fcc-Cs at the volume $V/V_0=0.44$. The squares are
the calculated results using the density-functional linear-response method and
the solid lines represent the results from a quasiharmonic calculation (see text).
The dash lines are the fitted results by the Born-von K\'{a}rm\'{a}n model. 
Also plotted in the right panel is the phonon
density of states derived from the linear-response calculations.\\

\noindent Figure 2: (a) The interplanar force constants (up to $n=8$) for
the $T_{[1\bar{1}0]}$[$\xi\xi$0] branch phonons of fcc-Cs at the
volume $V/V_0=0.44$. A clear oscillating behaviour against the number
of the plane (interplanar spacing) is seen. (b) Values of $\nu^2$
for the same branch phonons plotted as a function of the reduced wave-vector. 
The fitted curve (solid line) with 8 planes are shown together with the
contributions from the first five Fourier components. \\

\noindent Figure 3: Calculated phonon frequencies $\nu_i$ (THz) at the
high-symmetry {\sl L, X, K} and {\sl W} points for fcc Cs as a function 
of the volume. (a) Longitudinal and (b) transverse branches. \\

\noindent Figure 4: Calculated $T_{[1\bar{1}0]}$[$\xi\xi$0]
phonon-dispersion curve for fcc Cs at the volumes $V/V_0=0.37, 0.40,
0.41$, and $0.44$. The points are the calculated values and the lines 
result from the interpolation between points and are only a guide to the eye. \\

\noindent Figure 5: Schematic illustration of displacement pattern (left)
of atoms in fcc-Cs corresponding a $T_{[1\bar{1}0]}$[$\frac{1}{4}\frac{1}{4}$0] 
phonon and the resulting Cs superstructure (right). The lengths of solid arrows 
are proportional to the displacement of atoms and the dash arrows indicate the 
[100], [010] and [110] directions of fcc lattice.\\

\clearpage
\begin{figure}[h]
\includegraphics[angle=0,scale=0.9,bb= 50 10 554 774]{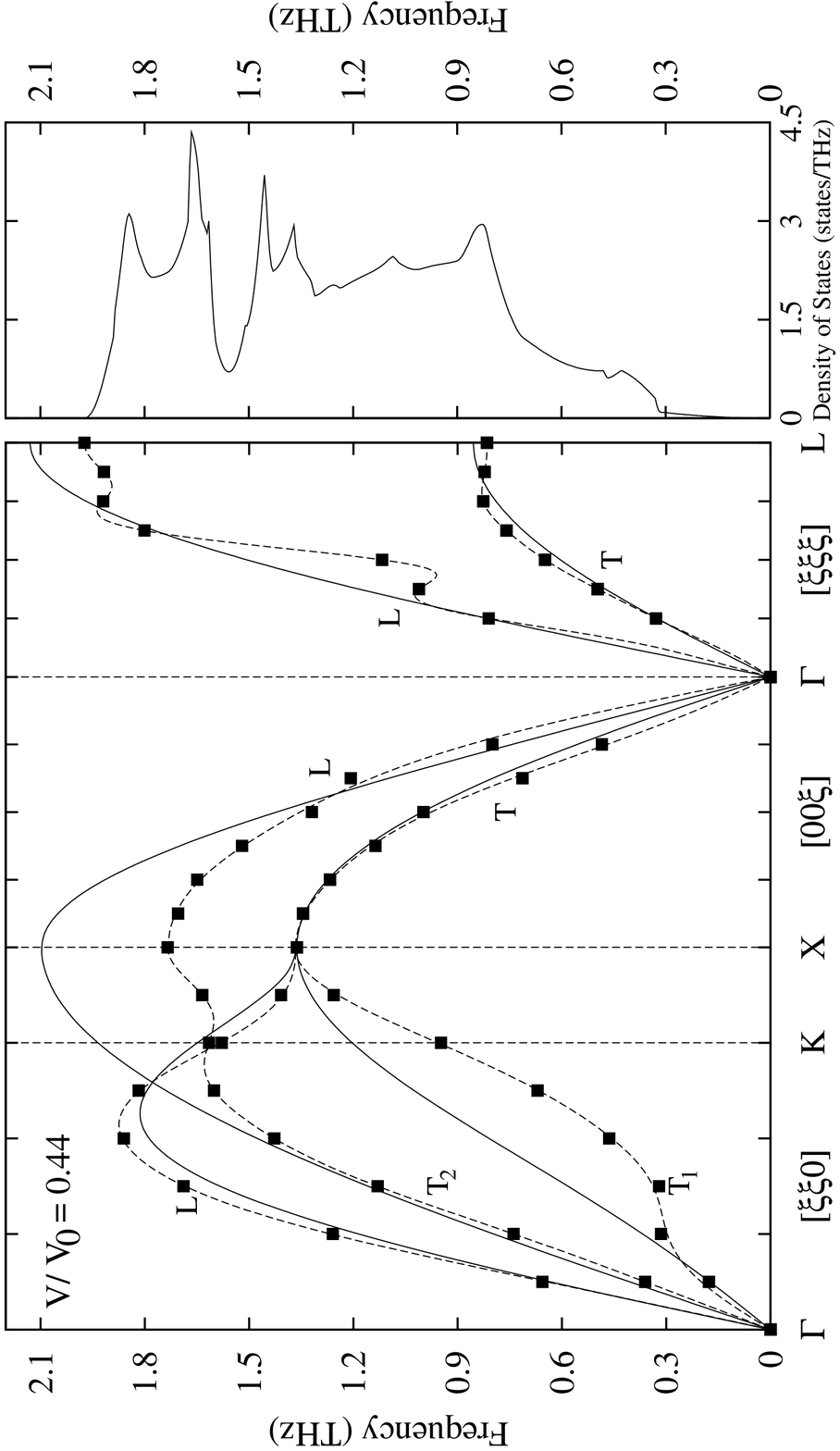}
\begin{center}\Large Fig. 1 of Y. Kong et al.\\
\end{center}
\end{figure}

\begin{figure}[h]
\includegraphics[angle=0,scale=1.0,bb= 50 50 534 736]{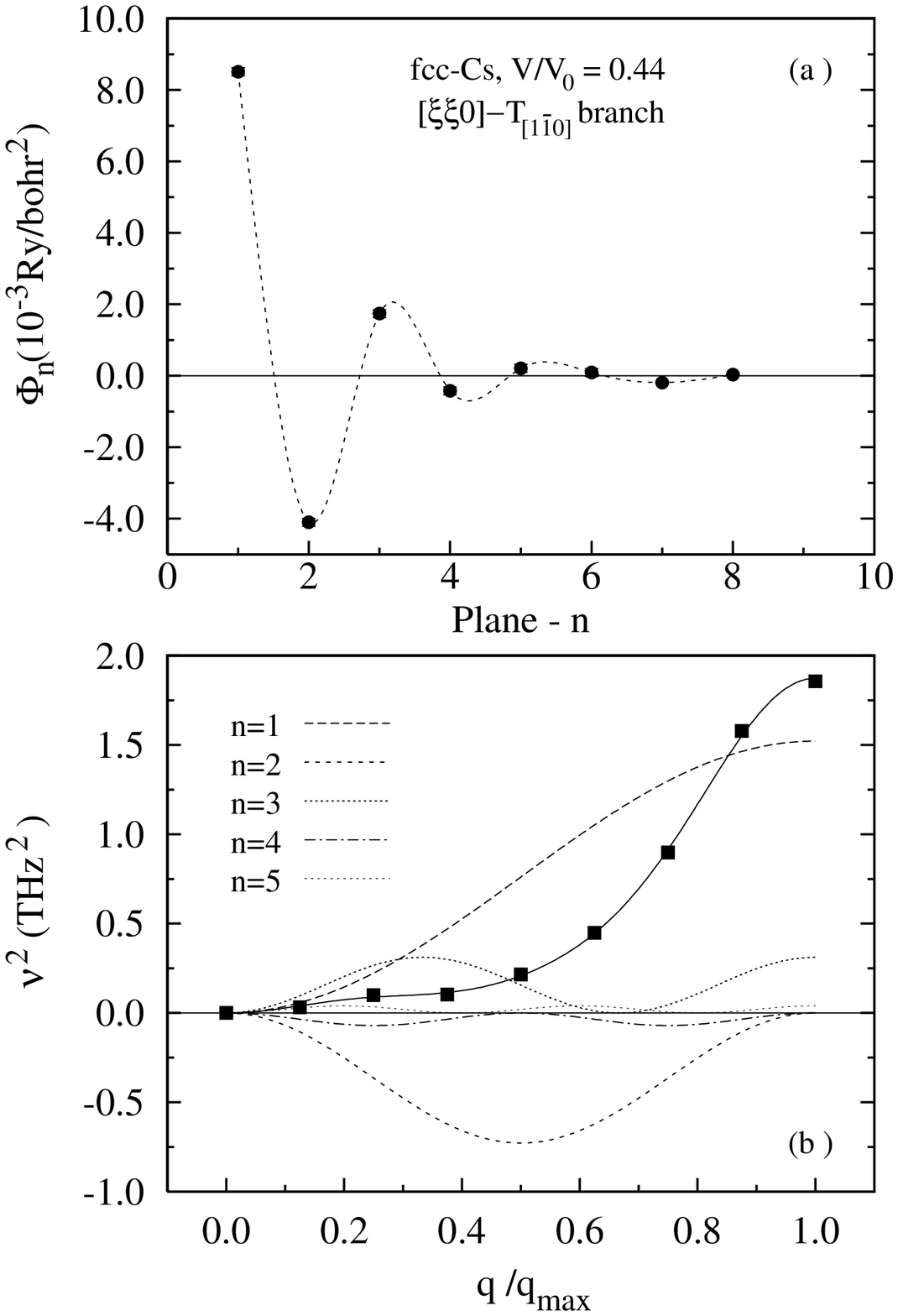}
\begin{center}\Large Fig. 2 of Y. Kong et al. \\
\end{center}
\end{figure}

\begin{figure}[h]
\includegraphics[angle=0,scale=0.9,bb= 45 60 594 746]{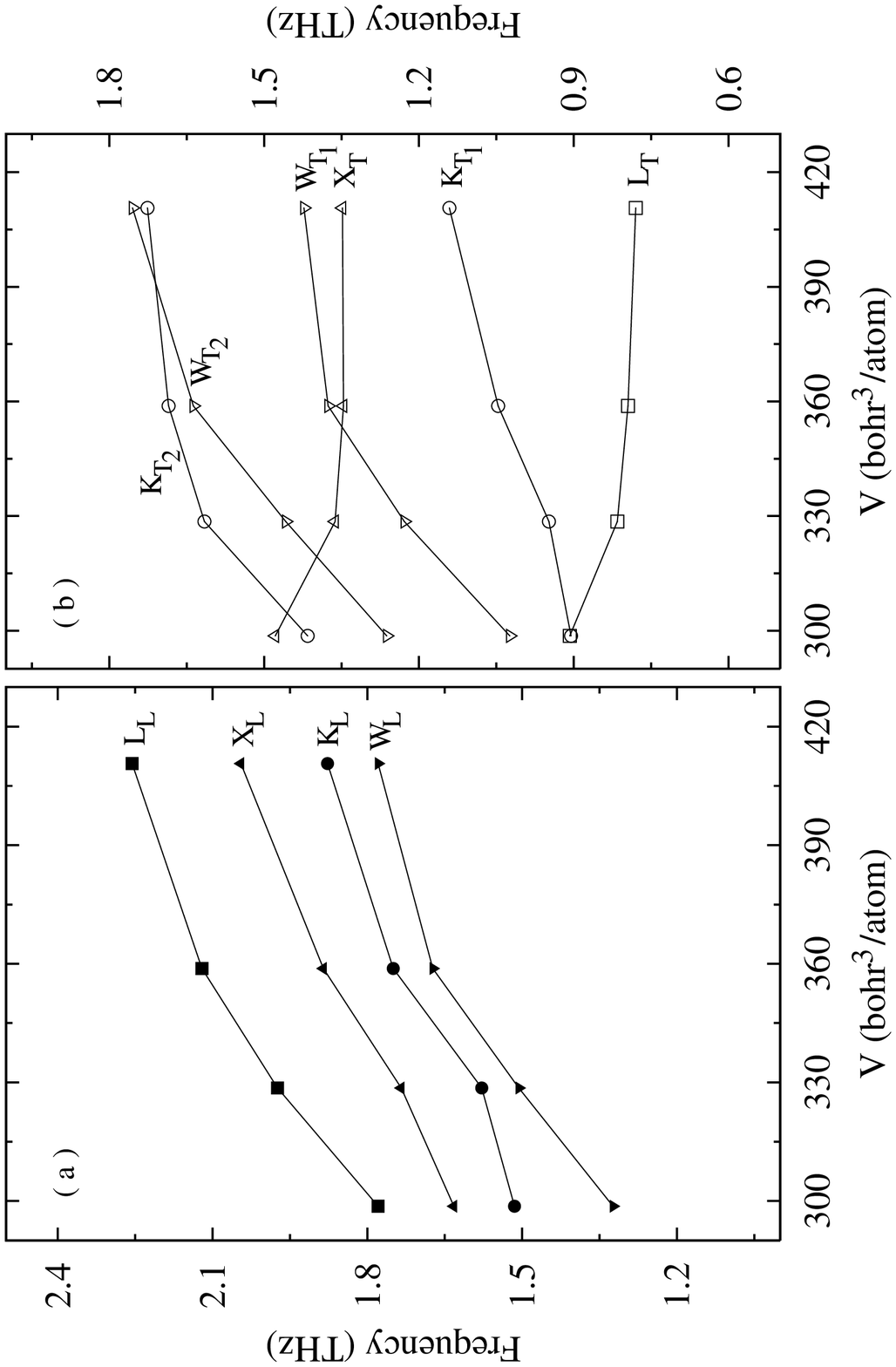}
\begin{center}\Large ~\\ ~\\ Fig. 3 of Y. Kong et al. \\
\end{center}
\end{figure}

\begin{figure}[h]
\includegraphics[angle=-90,scale=1.0,bb= 50 65 564 488]{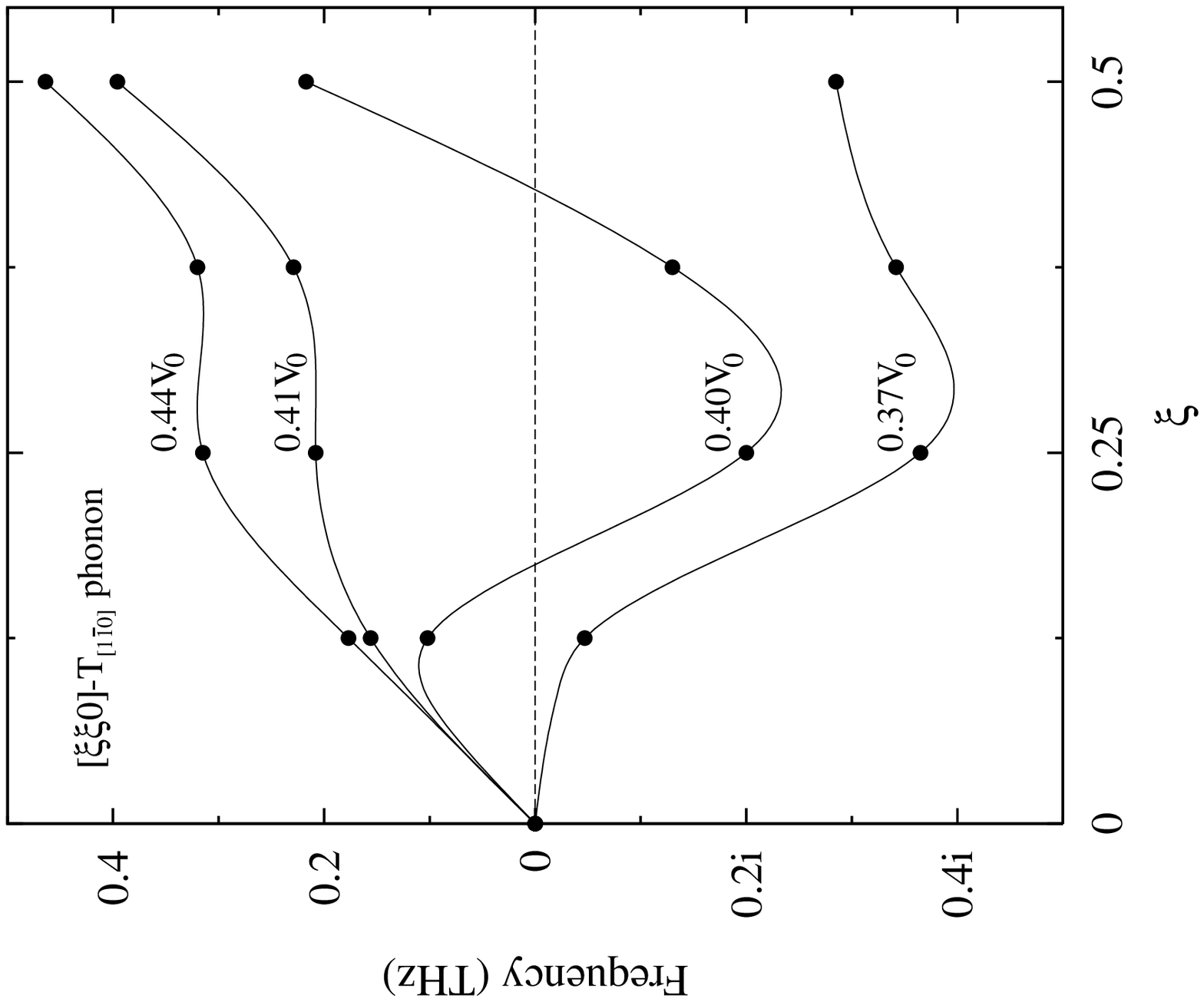}
\begin{center}\Large ~\\ ~\\ ~\\ ~\\ Fig. 4 of Y. Kong et al. \\
\end{center}
\end{figure}

\begin{figure}[h]
\includegraphics[angle=90,scale=0.75,bb= 62 60 790 461]{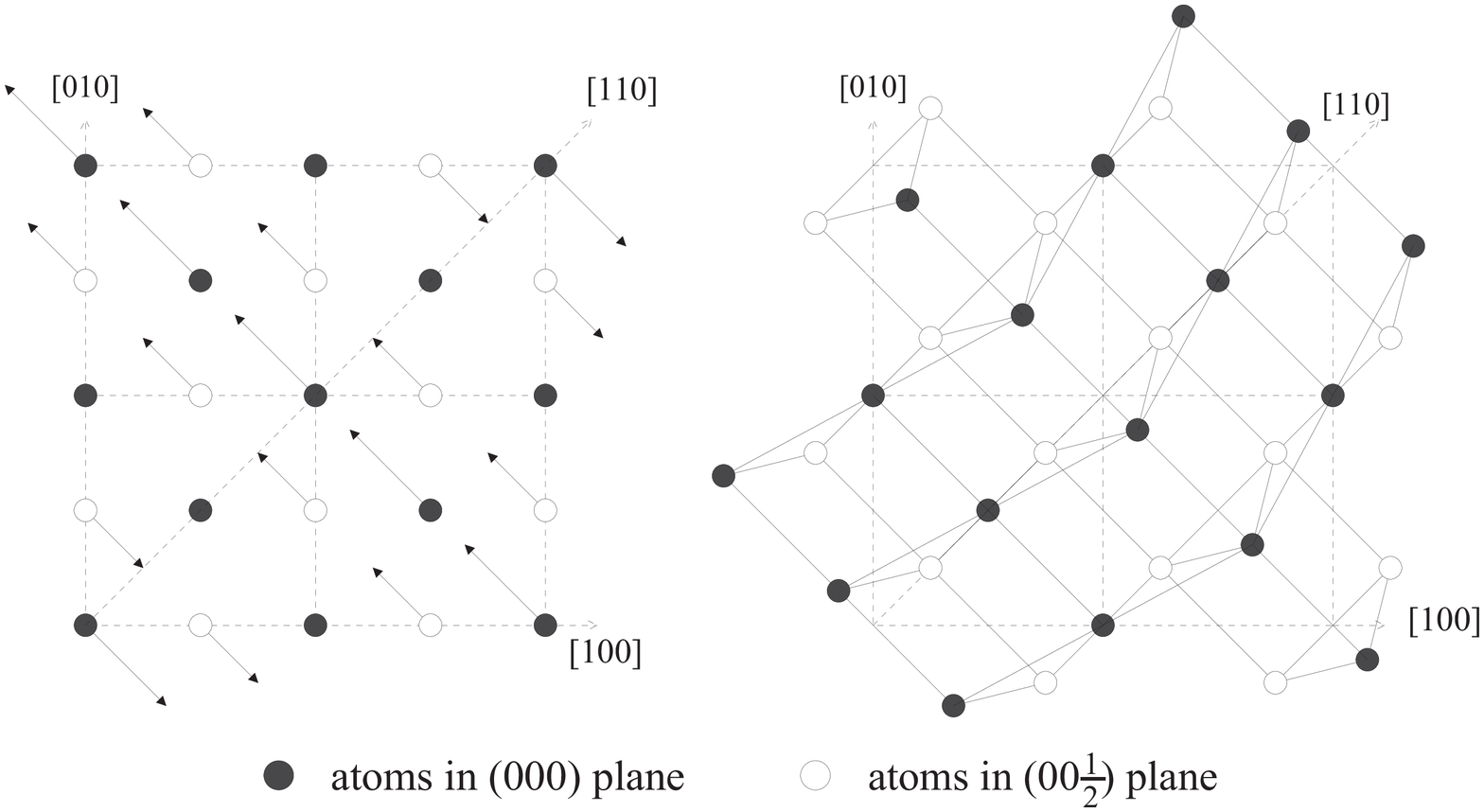}
\begin{center}\Large ~\\ ~\\ ~\\ ~\\ Fig. 5 of Y. Kong et al. \\
\end{center}
\end{figure}


\begin{thebibliography}{99}
\bibitem{cij}The bulk modulus and shear moduli ($C^{\prime}$, $C_{44}$)
of fcc-Cs are calculated in a full-potential LMTO scheme. For the volume
0.44$V_0$, $C_{11}$=13.8GPa, $C_{12}$=11.0GPa and $C_{44}$=8.1GPa.
\end{thebibliography}
\end{document}